\documentstyle[aps,preprint]{revtex}

\draft
\tighten

\def\Lie{\pounds}

\begin{document}
  \begin{flushright} \begin{small}
     gr-qc/0001088
  \end{small} \end{flushright}
\vspace{.5cm}
%%%%  Title  %%%%
\begin{center}
{\large \bf A Symplectic Hamiltonian Derivation of Quasilocal Energy-Momentum
            for GR}
\vskip.3cm
%%%%%  Authors  %%%%
Chiang-Mei Chen\footnote{E-mail: cmchen@joule.phy.ncu.edu.tw}
and James M. Nester\footnote{E-mail: nester@joule.phy.ncu.edu.tw}
\vskip.1cm
%%%%%  Address  %%%%
\smallskip
{\em Department of Physics and Center for Complex Systems,} \\
{\em National Central University, Chungli 320, Taiwan}
\vskip.1cm
%%%%%  Date  %%%%
{\small(Jan 26, 2000)}
\end{center}
%\maketitle

\begin{abstract}
The various roles of boundary terms in the gravitational Lagrangian
and Hamiltonian are explored.  A symplectic Hamiltonian-boundary-term
approach is ideally suited for a large class of quasilocal
energy-momentum expressions for general relativity.  This approach
provides a physical interpretation for many of the well-known
gravitational energy-momentum expressions including all of the
pseudotensors, associating each with unique boundary conditions.  From
this perspective we find that the pseudotensors of Einstein and M{\o}ller
(which is closely related to Komar's superpotential) are especially natural,
but the latter has certain shortcomings.  Among the infinite possibilities, we
found that there are really only two Hamiltonian-boundary-term
quasilocal expressions which correspond to {\em covariant} boundary
conditions; they are respectively of the Dirichlet or Neumann type.
Our Dirichlet expression coincides with the expression recently
obtained by Katz and coworkers using Noether arguments and a fixed
background.  A modification of their argument yields our Neumann
expression.

\end{abstract}


%\begin{multicols}{2} %%%%
%\narrowtext

%%%%%%%%%%%%%%%%%%%%%%%%%%%%%%%%%%%%%%%%%%%%%%%%%%%%%%%%%%%%%%%%%%%%%%
%                                                                    %
%%%%%%%%%%%%%%%%%%%%%%%%%%%%%%%%%%%%%%%%%%%%%%%%%%%%%%%%%%%%%%%%%%%%%%
\section{Introduction}
Via their energy-momentum density, material sources generate
gravitational fields.  Sources interact with the gravitational field
locally, hence they should be able exchange energy-momentum with the
gravitational field --- locally.  From this physical conception we are
led to expect the existence of a local density for gravitational
energy-momentum.  However the identification of a good localization
for gravitational energy-momentum has turned out to be an outstanding
fundamental problem.  Standard techniques led only to various
reference frame dependent complexes referred to as {\em
pseudotensors}.  This result can be understood in terms of the {\em
equivalence principle} which implies that one can not detect any
feature of the gravitational field at a point.  Consequently, the
whole idea has been criticized (see, e.g., \cite{MTW}, p 467) and the
pseudotensor approach in particular has largely been abandoned.

A new idea:  {\em quasilocal} (i.e., associated with a closed
2-surface) \cite{Pen82} has become widely accepted.  Many quasilocal
proposals have been considered (for the older works see \cite{BY93};
more recent works are cited in \cite{CN99,CNC99}).  Many well-known
quasilocal expressions obtained by different approaches have been
discussed in the literature; although they generally give different
values \cite{Ber92a}, most seem to work well enough at least for
certain physical situations.  A number of criteria for selecting a
good quasilocal expression (see, e.g., \cite{CY88}), including good
limits at spatial infinity, at future null infinity, to weak fields,
and to flat spacetime have been advocated.  Such requirements,
however, have proved to be insufficient; in fact it has been noted
that there still exist an infinite number of expressions satisfying
the proposed criteria \cite{Ber92b}.  We infer that additional {\em
principles} and {\em criteria} are very much needed to reduce and to
parameterize, if not entirely eliminate, the ambiguity.

One might hope that there would exist a ``best'' gravitational
energy-momentum expression which has either not yet been identified or
at least not yet accorded widespread acceptance.  On the other hand,
it is well to keep in mind that there are physical situations where
there is not one unique energy.  One example is thermodynamics,
wherein there are several energies (viz., internal, enthalpy, Gibbs
and Helmholtz) corresponding to different choices of boundary
conditions and independent variables; each one gives the relevant
value of the energy for a particular physical situation.  An even more
appropriate example is electrostatics.  It is well known that the work
done in moving a system of charges and dielectrics differs depending
on whether one holds the potential or the charge density fixed.  We
expect gravity to behave in a similar fashion:  consequently {\em
various definitions of gravitational energy may each be associated
with their own unique boundary condition.} Fortunately, there exists a
systematic technique, {\em symplectic analysis} \cite{KT79}, which,
along with its associated control-response relations, can be used to
identify the relationship between an energy-momentum expression and
the associated boundary conditions.

Our approach to quasilocal energy-momentum is by way of the
Hamiltonian formulation --- essentially, we take energy to be given by
the value of the Hamiltonian.  The rationale goes back to Noether's
work connecting symmetries and conserved currents:  in particular
energy-momentum is associated with translations in spacetime.  The
Hamiltonian is the Noether canonical generator of timelike
displacements.  The Hamiltonian for gravitating systems for a finite
region of spacetime includes, in addition to a volume density term, a
surface term which plays a key role.  Its value will determine the
quasilocal energy-momentum and, through its variation, it governs
the associated boundary conditions.  In an earlier work \cite{CN99} we
presented our ideas, applied to rather general geometric gravity
theories including Einstein's general relativity (GR), in terms of
differential forms.  However that technique is not so well known by
many people interested in gravitational energy-momentum; moreover most
people are mainly interested specifically in GR.  Hence we present
here a formulation of the application of our ideas to GR in the more
traditional holonomic (coordinate basis) style.  This will serve not
only to make our ideas more widely accessible but will also facilitate
comparison with the results obtained by other investigators, e.g.,
\cite{BFFV94,FF90,JS98,BY93,Tra62,PK99}.

Elsewhere we have used our Hamiltonian-boundary-term approach to
quasilocal quantities to show that all of the pseudotensors give well
defined quasilocal energy-momentum values, each of which is associated
with a particular choice of boundary conditions \cite{CNC99}.  From
the Hamiltonian boundary approach we find that two, the Einstein and
M{\o}ller expressions, each arise naturally, although the latter has a
few extra shortcomings.  A more important failing however, in our
opinion, is that none of the pseudotensors is associated with a
truly {\em covariant} boundary condition.

For the required new {\em principle} and {\em criteria} to restrict
the GR quasilocal energy-momentum expression we have advocated the
{\em Hamiltonian boundary variation principle} and the criteria of
{\em covariance}.  For GR we found that there are only two {\em
covariant} quasilocal expressions (both of which depend on the choice
of a reference configuration on the boundary) which correspond,
respectively, to Dirichlet and Neumann boundary conditions.  At first
we were surprised to learn that our Dirichlet expression coincides
with an expression developed by Katz and coworkers
\cite{LBKB95,KBLB97,KL97} using a different approach based on a
Noether type argument at the Lagrangian level along with a fixed
global background geometry.  With hindsight we now see that this
agreement is related to the close connection between the Hamiltonian
and Noether translation current.  Here we show how to modify their
argument to also obtain our Neumann quasilocal energy-momentum
expression.  Along the way we clarify the roles of the boundary (or
total derivative) terms in the Lagrangian, the Hamiltonian and their
respective variations.

%%%%%%%%%%%%%%%%%%%%%%%%%%%%%%%%%%%%%%%%%%%%%%%%%%%%%%%%%%%%%%%%%%%%%%
%                                                                    %
%%%%%%%%%%%%%%%%%%%%%%%%%%%%%%%%%%%%%%%%%%%%%%%%%%%%%%%%%%%%%%%%%%%%%%
\section{The Symplectic Idea in General Relativity}
In this section we outline the symplectic idea for Lagrangian and
Hamiltonian formulations (here and elsewhere we are much influenced by
Kijowski and coworkers \cite{JK90,Kij85,Kij97,KT79}) for general
relativity (GR), Einstein's theory of gravity (a detailed discussion
for general geometric gravity theories, in terms of differential
forms, appears in \cite{CN99}).  The simple and direct way to reveal
the symplectic structure of a physical configuration is through the
variation of the associated Lagrangian or Hamiltonian.

\subsection{Lagrangian formulation}
Let us first briefly review some features of the Lagrangian
variational
principle for classical field theories.  For our purposes we found it
convenient to use the first order formalism.  For a field $\phi^A$
the first order Lagrangian scalar density has the form
\begin{equation}
{\cal L}=P_A^\mu\partial_\mu\phi^A-\Lambda(\phi^A,P^\mu_A).
\label{fol}
\end{equation}
The field equations (which in this formulation contain only first
derivatives of the fields)
are taken to be the Euler-Lagrange expressions implicitly determined
by the variation:
\begin{equation}
\delta{\cal L}=\partial_\mu (P^\mu_A \delta\phi^A)
+{\delta{\cal L}\over\delta\phi^A}\delta\phi^A+
{\delta{\cal L}\over\delta P_A^\mu}\delta P_A^\mu.
\label{variation}
\end{equation}
The action is obtained by integrating the Lagrangian density over a
spacetime region; the variation of the action is given by the integral
of (\ref{variation}).  When integrated over a spacetime region the total
derivative term becomes a boundary term.  Technically the variational
derivatives of the action are well defined only if the boundary term in
the variation vanishes.   That requirement shows what
must necessarily be held fixed on the boundary, this quantity is referred
to as the ``control variable''. In this case it is the field
$\phi^A$, thus this Lagrangian is differentiable only on the space of
fields for which $\phi$ is given a predetermined dependence on the
boundary.   The variation boundary term, moreover, has a certain
{\em symplectic} form which connects the ``control variable'' with an
associated ``response variable'' (in this case $P^\mu_A$).

We are only concerned with actions which do not depend on the
position except via the fields.  Hence they have an invariance under
local infinitesimal translations (i.e, diffeomorphisms)  --- which can
be represented
by Lie derivatives.  Thus, for an arbitrary vector field $N$,
\begin{equation}
\Lie_N{\cal L}:=\partial_\nu(N^\nu{\cal L})\equiv\partial_\mu (P^\mu_A
\Lie_N\phi^A) +{\delta{\cal L}\over\delta\phi^A}\Lie_N\phi^A+
{\delta{\cal L}\over\delta P_A^\mu}\Lie_N P_A^\mu.
\label{translation}
\end{equation}
From this identity we conclude, by taking $N^\mu$ to have
constant coefficients, that the {\em canonical energy-momentum}
density,
\begin{equation}
T^\mu{}_\nu:=P^\mu_A\partial_\nu\phi^A-\delta^\mu_\nu{\cal L},
\label{cemd}
\end{equation}
is conserved.  More precisely its divergence is proportional to a
combination of the field equations and hence vanishes ``on shell''.
Note that the canonical energy-momentum density is not unique in the
sense that we can add to it an expression which is automatically
divergence free.  Such an expression is necessarily of the form
$\partial_\gamma U_\nu{}^{\mu\gamma}$ where $U_\nu{}^{\mu\gamma}\equiv
-U_\nu{}^{\gamma\mu}$.  This ambiguity allows one to adjust the zero
of energy and has been exploited to find ``improved'' energy-momentum
tensors such as the symmetrized one constructed by Belinfante
\cite{Bel39} and Rosenfeld \cite{Ros40}.  On the other hand, since we
have also assumed invariance with non-constant $N^\mu$, we are also
requiring that (\ref{translation}) be identically satisfied for the
terms proportional to $\partial N$ (this is only possible if the list
of dynamic variables includes certain geometric variables).  In this
way we discover that $T^\mu{}_\nu$ itself is linear in the field
equations and thus vanishes ``on shell''.  In other words the
`conservation law' is actually a differential identity connecting the
field equations showing that they are not all independent, hence the
evolution of the dynamical variables is underdetermined---a fact which
is directly related to the local `translational' gauge (i.e.,
diffeomorphism) freedom of the theory.

Now let us apply this analysis to gravity.  There are several choices
of variables and Lagrangians which can be used.  Since we favor a
first order approach, a natural geometric choice is to regard the
metric and connection as independent fields.  Even within this overall
approach there are various options, in particular the metric degrees
of freedom can alternately be encoded in terms of an orthonormal frame
while the torsion free and metric compatibility conditions can be
imposed {\it a priori}, or enforced via Lagrange
multipliers, or they can be obtained as dynamic field equations (this
is easily done in the vacuum case which is all that we consider here).
All of these approaches merit consideration.  We have investigated
many of the possible combinations; our preliminary conclusion is that
they lead to essentially the same result \cite{changthes}.  Here we
consider explicitly only one case which is relatively simple in the
holonomic treatment.

The field equations of (vacuum) GR can be
obtained from a first order variational principle using the Hilbert
Lagrangian density in the so-called `Palatini' form \cite{Wal84}
\begin{equation}
{\cal L}_H := \pi^{\mu\nu} R_{\mu\nu}(\Gamma). \label{LH}
\end{equation}
Here we are using the contravariant metric density, defined by
$\pi^{\mu\nu} := (2\kappa)^{-1}\sqrt{-g} g^{\mu\nu}$, where
$\kappa:=8\pi G/c^4$, and the (symmetric, i.e.,
torsion free) connection coefficients, $\Gamma^\mu{}_{\alpha\beta}$, as
independent variables (while our
conventions are generally those of \cite{MTW},
our treatment can be compared with
\cite{Kij85,Kij97} in which the same variable combinations appear).
The variation of the Lagrangian density, after the usual integration
by parts, has the form
\begin{equation}
\delta {\cal L}_H = %\hbox{ (field eq. terms) }
      {\delta {\cal L}\over \delta \pi} \delta\pi
    + {\delta{\cal L}\over\delta\Gamma}\delta\Gamma
    + \partial_\gamma \left( \pi^{\beta\nu}\delta^{\gamma\alpha}_{\mu\nu}\;
      \delta \Gamma^\mu{}_{\alpha\beta} \right). \label{VH}
\end{equation}
The variational derivatives will give the desired field equations:
$R_{\mu\nu}=0$,
from the variation with respect to $\pi^{\mu\nu}$, and $D_\lambda
\pi^{\mu\nu}=0$ (equivalent to $D_\lambda g_{\mu\nu}=0$), from
the variation with
respect to $\Gamma$.
When integrated over a spacetime region, the total derivative term
gives rise to a boundary term.  This boundary term shows that the
control variable is the connection and the response variable is linear
in $\pi^{\mu\nu}$.  The variational derivatives are well
defined only if this boundary variation term vanishes \cite{RT74}.
For a finite region this means we must `control' or `hold fixed'
(i.e., give as a prespecified function)
$\Gamma$ on the boundary.
For an asymptotically flat region the connection vanishes
asymptotically, nevertheless $\Gamma=0$ is not a sufficient
boundary condition.  Since we must allow for
variations with the generic spatial fall offs $\delta \pi \sim
O(1/r)$, $\delta \Gamma \sim O(1/r^2)$,
the Lagrangian boundary variation term will yield a finite result in
the asymptotic limit.  Formally the situation is then described by
saying that, in this case, the variational derivatives of the action
are not well defined on the full space of asymptotically flat metric
and connection fields, but rather only on the subspace where we
actually fix the specific asymptotic form of $\Gamma$.  This `problem'
is closely related to the fact that the Hilbert Lagrangian density is
asymptotically $O(1/r^3)$; consequently the action diverges for
$r\to\infty$.  The remedy is simple:  adjust the Lagrangian density by
adding a total derivative term.

For GR, an obvious alternative is the ``first order'' (in derivatives
of the metric) Lagrangian density, initially introduced by Einstein,
which can be easy obtained by adding a total derivative term to
(\ref{LH}):
\begin{eqnarray}
{\cal L}_E &:=&
    {\cal L}_H + \partial_\gamma \left(
      \pi^{\beta\nu} \Gamma^\mu{}_{\alpha\beta}
      \delta^{\alpha\gamma}_{\mu\nu} \right) \nonumber\\
   &\equiv& \pi^{\mu\nu} \left(
      \Gamma^\alpha{}_{\mu\beta} \Gamma^\beta{}_{\nu\alpha}
    - \Gamma^\alpha{}_{\mu\nu} \Gamma^\beta{}_{\alpha\beta} \right).
      \label{LE}
\end{eqnarray}
The variation of the Einstein Lagrangian,
\begin{equation}
\delta {\cal L}_E = \hbox{ (fields eq. terms) }
    + \partial_\gamma \left( \Gamma^\mu{}_{\alpha\beta}
      \delta^{\alpha\gamma}_{\mu\nu} \;
      \delta \pi^{\beta\nu} \right),
\end{equation}
has the same field equation terms as before but a different boundary
term which reflects an alternate symplectic structure and shows that
the variable to be held fixed --- the ``control variable''--- is now
the contravariant metric density while the ``response variable'' is a
certain combination of the connection.  Now, for asymptotically flat
fall offs, the Lagrangian boundary variation term does vanish
asymptotically, so the variational derivatives are well defined on the
space of all asymptotically flat fields (a related fact is that the
Einstein action is finite).

However the big drawback is that we now have a response expression
which is linear in $\Gamma$, a non-tensorial, reference frame
dependent object (along with this the Lagrangian density itself is not
covariant).  The cure for this new `problem' is to introduce a
background (or reference) connection $\bar \Gamma$ (actually this is
really only essential on the boundary) and define
$\Delta\Gamma:=\Gamma-\bar{\Gamma}$.
The latter, being the difference between two connections, is a {\em
covariant} quantity, which can be used in the Lagrangian density
boundary (i.e., total derivative) term.  The `improved Einstein'
action is now
\begin{equation}
{\cal L}_{IE} =
    {\cal L}_H + \partial_\gamma \left(
      \pi^{\beta\nu} \Delta \Gamma^\mu{}_{\alpha\beta}
      \delta^{\alpha\gamma}_{\mu\nu} \right).
\label{IE}
\end{equation}
The variation gives
the same field equation terms but now has a {\em
covariant}
boundary-variation symplectic structure:
 \begin{equation}
 \delta {\cal L}_{IE} = \hbox{ (fields eq.  terms) }
    + \partial_\gamma \left(\Delta \Gamma^\mu{}_{\alpha\beta}
      \delta^{\alpha\gamma}_{\mu\nu} \;
      \delta \pi^{\beta\nu} \right).
 \end{equation}

The Lagrangian boundary term does not, as is well known, affect the
field equations.  What it does affect is the boundary conditions
implicit in the action.  From another point of view, changing the
action by a total derivative term amounts to a {\em canonical
transformation}; in particular, as we saw in the cases considered, it
is possible to interchange the role of `coordinate' and `momentum'.
Thus the Lagrangian variational boundary term possesses important
information:  the {\sl symplectic structure} representing the
control--response relation of the system \cite{KT79,Wal93}.  For
instance, the symplectic structure in (\ref{VH}) shows that the
connection is the control variable and the response is a certain
combination of the metric.

\subsection{Hamiltonian formulation}
The energy of a gravitating system can be identified with the value of
the Hamiltonian.  However the Hamiltonian approach necessitates a
splitting of spacetime at least to some extent.  One constructs a
$3+1$ foliation of spacetime by selecting a time function $t$ such
that the hypersurfaces, $\Sigma_t$, of constant $t$ are space-like
Cauchy surfaces.  The standard Hamiltonian formulation for general
relativity is the ADM representation (see.e.g., \cite{ADM,IN} and
\cite{MTW} Ch 21), in which 4-covariant objects are decomposed into
various 3-covariant parts.  In particular, the spacetime metric,
$g_{\mu\nu}$, is decomposed into the form,
\begin{eqnarray}
ds^2 &=& g_{\mu\nu} dx^\mu dx^\nu \nonumber\\
     &=& - N^2 dt^2 + h_{ab} (dx^a + N^a dt)(dx^b + N^b dt),
\end{eqnarray}
which depends on three spatially covariant parts:
 the {\sl lapse function} $N$, the {\sl shift
vector} $N^a$ and the spatial metric $h_{ab}$, induced on $\Sigma_t$.
The associated Hamiltonian density is
obtained from
\begin{equation}
{\cal H} = \frac{\partial {\cal L}}{\partial \dot h_{ab}} \dot h_{ab}
    - {\cal L},
\end{equation}
where $\dot h_{ab}\equiv \partial_t h_{ab} := \Lie_t h_{ab}$.
Although this approach has led to much insight, it has the
drawback that the resultant Hamiltonian
formulation is only manifestly 3-dimensionally covariant.

We prefer to use a more ``covariant'' approach to the Hamiltonian
density.  To this end it is convenient to use the so-called `Palatini'
method of treating the metric and connection independently.  (Our
approach is in many ways similar to that of Kijowski, see, e.g.
\cite{Kij97}.)\ Let us note some general features.  First of all,
since we are concerned with localization, we shall want to find the
Hamiltonian which can evolve a {\em finite} spatial region.  To
achieve our `covariant' formulation, we represent the time evolution
direction as a covariant 4-vector field $N^\mu$.  Quite generally the
(4-covariant) Hamiltonian --- which is essentially the Noether
generator of translations (i.e., Lie derivatives) along $N^\mu$ --- is
given by the spatial integral over (the finite spatial hypersurface)
$\Sigma_t$ of a 4-covariant Hamiltonian density.  In order to generate
the Lie derivatives, the Hamiltonian density is necessarily linear in
the time displacement vector field $N^\mu$ and its derivatives.
Consequently it can be expanded into the form
\begin{equation}
{\cal H}^\mu(N)=N^\nu {\cal H}^\mu{}_\nu + \partial_\nu [{\cal
B}^{\mu\nu}(N)],
\label{hamdecomp}
\end{equation}
where, it turns out that (at least for our representation) ${\cal
B}^{\mu\nu}(N)\equiv-{\cal B}^{\nu\mu}(N)$.

On the other hand, beginning from the Lagrangian density, we can apply
our Noether type argument to a translation along $N^\mu$,
(see \cite{CN99}).  Formally we then arrive at a conserved quantity
(essentially the canonical energy-momentum density discussed earlier) which
is actually just this same Hamiltonian density.
 From this analysis we learn a couple of
important things.  First, we find that (``on shell'') the Hamiltonian
density is necessarily conserved: $\partial_\mu
{\cal H}^\mu(N)\equiv0$.
At this point we want to call attention to the fact that
the possibility of adjusting the canonical energy-momentum density
(\ref{cemd})
by adding an automatically divergence free part exactly corresponds to
adjusting ${\cal B}^{\mu\nu}$ in (\ref{hamdecomp}).  Second, we find
that ${\cal H}^\mu{}_\nu$
is linear in the field equations and thus vanishes ``on shell''.  From
this latter fact we
conclude that the numerical {\em value} of the Hamiltonian is completely
determined by the $\partial {\cal B}$ term, which, when integrated over a
spatial region $\Sigma_t$, via the divergence theorem, gives rise to an
integral of  ${\cal B}$
over the 2-dimensional spatial boundary $\partial \Sigma_t$.
The {\em value} of the Hamiltonian for a finite region is thus
determined by the Hamiltonian boundary
term and hence is {\em quasilocal}: it is associated with the closed
spatial 2-surface boundary of the region.

Now we turn to specific details for GR.
Because we work with first order Lagrangians, we can easily obtain the
Hamiltonian by
merely rearranging the Lagrangian into the field theory analogue of $L=p_k
{\dot q}{}^k-H$; essentially from (\ref{fol}) we simply get ${\cal
L}=P_A^0\partial_t\phi^A-(-P^a_A\partial_a\phi^A+\Lambda)$.
We first consider the Hilbert Lagrangian which is given by
the spatial integral of the Hilbert Lagrangian density (\ref{LH}).
The spatial integrand can be expanded in the form
\begin{equation}
{\cal L}_H N^\mu d \Sigma_\mu = \left\{ \dot \Gamma^\alpha{}_{\beta\nu}
      \pi^{\beta\gamma} \delta^{\mu\nu}_{\alpha\gamma}
    - {\cal H}_H^{\mu}(N) \right\} d\Sigma_\mu,
\end{equation}
where $d\Sigma_\mu := \frac1{3!}
\epsilon_{\mu\alpha\beta\gamma} dx^\alpha dx^\beta dx^\gamma$
and our definition of
$\dot
\Gamma^{\alpha}{}_{\beta\gamma}$, which simply reduces to
$N^\mu \partial_\mu \Gamma^{\alpha}{}_{\beta\gamma}
=  \partial_t \Gamma^{\alpha}{}_{\beta\gamma}$ in adapted coordinates,
is given in general in
appendix \ref{dynamic} along with other
details regarding our choice of representation.
From (\ref{LH}), after a straightforward calculation,
without discarding any total derivative term,
we obtained the explicit expressions
\begin{eqnarray}
{\cal H}^\mu{}_\nu &=&
     -\frac12 R^\alpha{}_{\beta\gamma\lambda} \pi^{\mu\beta}
      \delta^{\mu\lambda\gamma}_{\alpha\nu\sigma}
    - \Gamma^{\alpha}{}_{\beta\nu} D_\gamma \pi^{\beta\lambda}
      \delta^{\mu\gamma}_{\alpha\lambda}, \label{hamden}\\
{\cal B}_H^{\mu\nu}(N) &=& N^\gamma \pi^{\beta\lambda}
      \Gamma^{\alpha}{}_{\beta\gamma} \delta^{\mu\nu}_{\alpha\lambda}.
\label{BH}
\end{eqnarray}
Note that, as predicted, the (spatial hyper-) surface density part ${\cal
H}^\mu{}_\nu$ vanishes `on shell', since it is
linear in the {\em non-metricity},
$D_\alpha g_{\mu\nu}$, and, with vanishing non-metricity,
the curvature
contractions reduce to the Einstein tensor, $G_{\mu\nu}$.
Hence, as expected, the value of the Hamiltonian
comes only from the boundary term ${\cal B}$, which gives the quasilocal
energy-momentum.

The Hamiltonian symplectic structure can be found, as in the
Lagrangian case, by varying the Hamiltonian (regarding it as a function of
$\Gamma^{\alpha}{}_{\beta\nu}$ and its conjugate momentum
$\pi^{\gamma(\beta}\delta^{\nu)\mu}_{\gamma\alpha}=
\delta^\mu_\alpha\pi^{\beta\nu}-\delta^{(\nu}_\alpha\pi^{\beta)\mu}$).
 This variation fits the general pattern \begin{equation}
\delta {\cal H}^\mu(N) = \hbox{ (field equation terms) }
    + \partial_\nu {\cal C}^{\mu\nu}(N).
\end{equation}
The field equation terms include a set of initial value constraints
and dynamical equations \cite{MTW,ADM,IN} which may be used to
calculate the evolution of the gravitational fields.  Here our focus,
however, is on the variational boundary term ${\cal C}^{\mu\nu}=-{\cal
C}^{\nu\mu}$ which reflects the symplectic structure --- and the
implicitly built in boundary conditions --- of the physical system
with respect to the particular Hamiltonian density under
consideration.  The variation of the spatial hypersurface part,
$N^\mu{\cal H}^\nu{}_\mu$, in addition to the field equation terms,
gives rise to the total divergence
\begin{equation}
\partial_\tau[N^\lambda(\delta\Gamma^\alpha{}_{\beta\mu}\pi^{\beta\sigma}
\delta^{\tau\rho\mu}_{\alpha\sigma\lambda}
-\Gamma^\alpha{}_{\beta\lambda}\delta\pi^{\beta\sigma}
\delta^{\tau\rho}_{\alpha\sigma})].
\label{varhamden}
\end{equation}
Combining this with the variation of the boundary term (\ref{BH}), we
find that, for the present (Hilbert) case
the Hamiltonian variational boundary term, ${\cal C}$, takes the
explicit form \begin{equation}
{\cal C}_H^{\tau\rho}(N) =
    - 2 \pi^{\beta\nu} \,       \delta \Gamma^\mu{}_{\alpha\beta}\,
\delta^{\alpha[\tau}_{\mu\nu} N^{\rho]},
\end{equation}
showing that the `control variable' is --- similar to the Lagrangian
case --- (certain projected components of) the connection
coefficients.

Expanding out the Hamiltonian boundary expression (\ref{BH}), and
using the $Dg=0$ field equation to express the connection coefficients
in terms of the metric gives
 \begin{equation}
{\cal B}_H^{\mu\nu}(N) \equiv \kappa^{-1}\sqrt{-g} N^\gamma
 g^{\beta[\nu}\Gamma^{\mu]}{}_{\beta\gamma} \equiv
(2\kappa)^{-1}\sqrt{-g}N^\gamma g^{\beta\nu} g^{\mu\sigma}
(\partial_\beta g_{\sigma\gamma}-
\partial_\sigma g_{\beta\gamma}).
\label{bh}
\end{equation}
This is in fact the {\em superpotential} which gives rise to the
M{\o}ller {\em pseudotensor} \cite{Mol58}.  From this calculation we
have acquired two insights:  first, the M{\o}ller pseudotensor is
essentially a {\em quasilocal} object and, second, it really gives the
{\em energy} --- the value of the Hamiltonian --- for the particular
Hamiltonian which generates time displacements in the case that the
connection is fixed on the boundary.

However, there are some shortcomings in this Hamiltonian (aside from
the obvious fact that the boundary term is not covariant, which we
will remedy further below).  First, although the boundary condition at
infinity is simply $\Gamma=0$, we must consider the rate of approach.
With the standard fall offs, in particular $\delta\Gamma\to O(1/r^2)$,
the boundary term in the variation of the Hilbert Hamiltonian will not
automatically vanish asymptotically, indicating that the Hamiltonian
is not differentiable on the phase space of all asymptotically flat
fields; hence one must actually give the explicit asymptotic
functional form for $\Gamma$ at each instant of time.  Second, the
actual value of the energy calculated from the boundary term for the
Schwartzschild solution {\em is not} the expected value $M$ but
exactly {\em half} of that amount.

This fact is closely connected with a well-known problematical
feature of Komar's covariant expression\cite{Kom},
\begin{equation}
{\cal
B}^{\tau\rho}_K(N):={1\over\kappa}\sqrt{-g}D^{[\tau}N^{\rho]}
={1\over\kappa}\sqrt{-g}\big(g^{\mu[\tau}\partial_\mu N^{\rho]}
+N^\gamma g^{\mu[\tau}\Gamma^{\rho]}{}_{\gamma\mu}\big),
\label{komar}
\end{equation}
which is equivalent to the M{\o}ller superpotential (\ref{bh}) when
the components of $N^\mu$ are constant.  Long ago it was noted that if
the Komar expression is normalized to give the correct energy-momentum
then it gives a value for the angular momentum which is twice that
desired, conversely if it is normalized to give the correct angular
momentum it then gives only half of desired amount for the energy.
(The proper way to reconcile the M{\o}ller-Komar superpotential with
the desired energy-momentum and angular momentum results was found
some time ago \cite{Katz,Sor,Chr}.  In fact the results of those works
are forerunners of our preferred expressions discussed below.) Here we
found that, from the standard normalization of the Hilbert Lagrangian,
the associated boundary term in the Hamiltonian (obtained without
discarding or modifying any boundary terms) naturally gives rise to
the M{\o}ller-Komar superpotential with the latter normalization.

M{\o}ller himself later noted that the Hilbert Lagrangian leads to his
superpotential \cite{Mol61}.  Also, it has long been known that the
Hilbert Lagrangian leads via a Noether argument to the Komar
superpotential (see, e.g., \cite{Bergman}).  (The Komar potential has
also been obtained in a Hamiltonian treatment via a Legendre
transformation from the Einstein Hamiltonian \cite{Chr}.)\ \ Although
the factor of 2 problem with Komar's expression has also long been
known, yet it seems that only very recent works \cite{IW94,CJMC98}
have explicitly noted that the normalization arising directly from the
Hilbert Lagrangian gives only half of the expected energy-momentum.

There is a very simple cure to this problem of getting half of the
desired value.  Exploiting the freedom in the Hamiltonian that we
noted above --- the freedom to add a divergence free term to the
canonical energy-momentum density without changing the fact that it is
conserved --- we can modify the Hamiltonian boundary term.  Adjusting
the Hamiltonian boundary term `by hand' will not change the equations
of motion but it will change the boundary conditions and the value of
the quasilocal energy-momentum.  Note that such an adjustment is
essential if we wish to obtain a Hamiltonian which will be
differentiable on the phase space of all asymptotically flat fields,
as was nicely explained some time ago \cite{RT74} in connection with
the usual ADM formulation.  (Indeed the usual approach is simply to
discard all boundary terms on the way from the Lagrangian to the
Hamiltonian and then to fix up the Hamiltonian boundary term in the
end to produce the desired behavior.) \ \ Given this freedom, we could
simply double the boundary term in the Hilbert Hamiltonian.  This
would certainly take care of the problem of getting only half the
value for the energy, {\em but} the symplectic structure in the
variation of the Hamiltonian would then be modified.  The necessary
boundary condition would then require the vanishing of
\begin{equation}
\partial_\tau \left\{ N^\lambda \left(
     \Gamma^\alpha{}_{\beta\lambda} \delta\pi^{\beta\sigma}
     \delta^{\rho\tau}_{\alpha\sigma}
   + \delta\Gamma^\alpha{}_{\beta\lambda} \pi^{\beta\sigma}
     \delta^{\tau\rho}_{\alpha\sigma}
   + 2 \delta\Gamma^\mu{}_{\alpha\beta} \pi^{\beta\nu}
     \delta^{\alpha[\tau}_{\mu\nu} \delta^{\rho]}_\lambda \right) \right\},
\end{equation}
which leads to a rather unattractive, complicated boundary condition
requiring the vanishing of a combination of $\delta \pi$ and
$\delta\Gamma$ and which, moreover, would not automatically vanish
asymptotically --- so the Hamiltonian would still not be
differentiable on the space of all asymptotically flat fields.

Let us now briefly consider the Einstein
Lagrangian density (\ref{LE}). We take the dynamical
``coordinate'' variable to be $\pi^{\beta\gamma}$.
The associated Hamiltonian density can be found from
\begin{equation}
{\cal L}_E N^\mu d \Sigma_\mu = \left\{ \dot \pi^{\sigma\alpha}
      \Gamma^\gamma{}_{\sigma\nu} \delta^{\mu\nu}_{\alpha\gamma}
    - {\cal H}_E^{\mu}(N) \right\} d\Sigma_\mu,
\end{equation}
(our general definition of $\dot \pi$ is given in appendix
\ref{dynamic}; in adapted coordinates it simply reduces to
 $\partial_t \pi^{\sigma\alpha}$).
The Hamiltonian density ${\cal H}_E$ still has the same ADM surface
part ${\cal H}^\mu{}_\nu$ (but when varied it is now to be regarded as
a function of
$\pi^{\sigma\alpha}$ and its conjugate momentum,
${1\over2}\Gamma^\gamma{}_{\gamma\sigma}\delta^\mu_{\alpha}
+{1\over2}\Gamma^\gamma{}_{\gamma\alpha}\delta^\mu_{\sigma}
-\Gamma^\mu_{\sigma\alpha}$).
 However the Hamiltonian has now acquired a different boundary term
given by
 \begin{equation} {\cal B}_E^{\mu\nu}(N) := N^\tau \pi^{\beta\lambda}
      \Gamma^\alpha{}_{\beta\gamma}
      \delta^{\mu\nu\gamma}_{\alpha\lambda\tau}.
\end{equation}
This Hamiltonian boundary term, which arose directly from the
Einstein Lagrangian density without discarding or adjusting any exact
differentials, is a familiar object.  Using the metric compatible
field equation to replace the connection by derivatives of the metric
leads to a well-known form of the expression,
\begin{equation}
{\cal B}_E^{\mu\nu}(N) \equiv (2\kappa)N^\lambda
(-g)^{-1/2}g_{\lambda\tau}\partial_\gamma
(\pi^{\mu\tau}\pi^{\nu\gamma}-
\pi^{\nu\tau}\pi^{\mu\gamma}).
\end{equation}
This is exactly the Freud superpotential \cite{Freud} whose divergence
gives rise to the Einstein pseudotensor.  The spatial integral of the
Einstein pseudotensor yields a value which is actually quasilocal, it
is given by the integral of the Freud superpotential over the closed
2-boundary of the spatial region.  This is identically the same
boundary integral and thus the same quasilocal value as is determined
by the Einstein Hamiltonian via its boundary term.  Extending the
region to spatial infinity yields the total energy-momentum, now with
the proper normalization.

The boundary term in the variation of the Einstein Hamiltonian
takes a form which differs from the Hilbert case:
  \begin{equation}
{\cal C}_E^{\tau\rho}(N) = 2 \Gamma^\mu{}_{\alpha\beta}      \,
      \delta \pi^{\beta\nu}\delta^{\alpha[\tau}_{\mu\nu} N^{\rho]} .
\label{BE}
\end{equation}
From the symplectic structure of this Hamiltonian variation boundary
term we learn that this Einstein choice corresponds to holding fixed
the contravariant metric density.  With the usual asymptotics this
term will vanish at spatial (but not at future null) infinity,
consequently the Einstein Hamiltonian is automatically differentiable
on the phase space of all asymptotically flat fields (spatially, while
at future null infinity one must specify the detailed functional
asymptotic form of the metric to describe the radiation).  The
symplectic response, however, reveals a deficiency.  Since it is some
projected components of the connection, it is not really a covariant
object.  An improved result could be obtained from the Lagrangian
density (\ref{IE}), but we have already seen the important ideas so,
instead of elaborating that case, we will just go on to our final
forms for the Hamiltonian boundary term in the next section.  However
before we do that let us make a few observations.

The role of the variational boundary term in the Lagrangian and
Hamiltonian formulations are similar, in that in both instances one
can adjust the boundary term to change the implicit boundary
conditions.  However in the Hamiltonian case there is an additional
entirely independent and very strong motivation which draws our
attention to the boundary term and moreover invites us to modify it.
Because the Hamiltonian is conserved, its value has a physical
significance not shared by the Lagrangian.  This conservation property
is preserved under modifications of the boundary term (equivalently,
preserved under adding a divergence free term to the Hamiltonian
density).  In the case of the Hamiltonian for dynamic geometry, the
entire value actually comes from the boundary term.

One important way in which the boundary terms in the Lagrangian and
Hamiltonian differ is that the former determines a boundary condition
on the whole boundary of a spacetime region whereas the latter
determines only spatial boundary conditions.  By adjusting both of
them accordingly we can independently choose what is held fixed on the
initial time hypersurface and at the spatial boundary (which in fact
is convenient for the different types---Cauchy vs.\
Dirichlet/Neumann---of boundary conditions typically required on these
surfaces).  This fact is related to another way in which the
Hamiltonian boundary term issue differs from that of the Lagrangian.
At the Hamiltonian level, we can make boundary terms which depend on
the displacement vector field $N^\mu$ in various ways.  This allows
for many more possibilities than those like (\ref{LE}) and (\ref{IE}) that are
available at the purely Lagrangian
level.  Consequently there is a bigger need for a suitably restrictive
criterion.

The plain fact is that we can entirely ignore the boundary term which
arises from the Lagrangian and simply change the Hamiltonian boundary
term to anything we want.  However our choice is constrained if we
wish to satisfy an important physical desiderata:  namely to get the
desired energy-momentum values for empty space, weak fields and at
spatial and future null infinity.  This requirement shows up only at
the Hamiltonian level; it is easily dealt with at that level
whereas in general it is not so readily satisfied
by a judicious adjustment of the boundary term back at the Lagrangian
level.  In fact  this requirement
{\em forces} us to adjust the Hamiltonian boundary term away from that
naturally inherited from the Hilbert Lagrangian.  Moreover, it actually
fixes the form of the Hamiltonian boundary term --- but only to linear
order.  Going beyond the linear order we can use our freedom to build
in certain boundary conditions via the Hamiltonian variation
symplectic structure.

 One consequence of this freedom
is that, not only the superpotentials for the M{\o}ller and Einstein
pseudotensors, but in fact also the superpotentials associated with
{\em all} of the other pseudotensors are likewise acceptable
Hamiltonian boundary terms.
Here we briefly outline the argument which we have presented in more
detail elsewhere \cite{CNC99}.  Consider the pseudotensor
idea: a suitable {\em superpotential}
 $H_\mu{}^{\nu\lambda}\equiv H_\mu{}^{[\nu\lambda]}$ is
selected and used to split the Einstein tensor thereby defining the
associated gravitational energy-momentum pseudotensor:
 \begin{equation}
\kappa \sqrt{-g} N^\mu t_\mu{}^\nu:=-N^\mu\sqrt{-g}G_\mu{}^\nu+
\frac{1}{2}\partial_\lambda( N^\mu H_\mu{}^{\nu\lambda}),
\end{equation}
where we have inserted a vector field to
make the calculation more nearly covariant.  The usual
formulation is recovered by taking the components of the vector field
to be constant in the present reference frame.   Einstein's
equation, $G_\mu{}^\nu=\kappa T_\mu{}^\nu$,
can now be rearranged into a form where the source is the {\em total}
effective energy-momentum pseudotensor
\begin{equation}
\partial_\lambda H_\mu{}^{\nu\lambda}=
2\kappa \sqrt{-g}{\cal T}_\mu{}^\nu:=
2\kappa \sqrt{-g}( t_\mu{}^\nu + T_\mu{}^\nu).
\end{equation}
An immediate consequence of the antisymmetry of the superpotential is
that ${\cal T}_\mu{}^\nu$ is a conserved
current: $\partial_\nu [(-g)^{1/2} {\cal T}_\mu{}^\nu]\equiv0$,
which integrates to give a conserved energy-momentum.
The energy-momentum within a finite region
\begin{eqnarray}
&&-P(N):=-
\int_{\Sigma} N^\mu {\cal T}_\mu{}^\nu \sqrt{-g}d\Sigma_\nu
\nonumber\\
&\equiv&\int_{\Sigma} \bigl[ N^\mu \sqrt{-g}(\frac{1}{\kappa}
G_\mu{}^\nu-T_\mu{}^\nu) - \frac{1}{2\kappa}
\partial_\lambda (N^\mu H_\mu{}^{\nu\lambda})\bigr]d\Sigma_\nu
 \nonumber\\
&\equiv&\int_{\Sigma} N^\mu {\cal H}^\nu{}_\mu d\Sigma_\nu +
 \oint_{S=\partial \Sigma}{\cal B}(N)\equiv H(N),
\label{Ham}
\end{eqnarray}
is seen to be just the value of the Hamiltonian.  Note that
${\cal H}^\nu{}_\mu$ is the covariant form of the ADM Hamiltonian
density, which has a vanishing numerical value, so that
the value of the Hamiltonian is determined purely by the boundary term
${\cal B}(N)=-N^\mu (1/2\kappa)H_\mu{}^{\nu\lambda}
(1/2)dS_{\nu\lambda}$.
Thus for any pseudotensor the associated {\em superpotential} is
naturally a Hamiltonian boundary term.  Moreover the energy-momentum
defined by such a pseudotensor does not really depend on the local
value of the reference frame, it is actually {\em quasilocal}---it
depends (through the superpotential) on the values of the
reference frame (and the fields) only on the boundary of a region.

 The Hamiltonian approach endows these quasilocal values with
a physical significance.  To understand the {\em physical meaning} of
the quasilocalization, calculate the boundary term in the Hamiltonian
variation:
\begin{equation}
-{1\over2}\left[\delta\Gamma^\alpha{}_{\beta\lambda}N^\mu
\pi^{\beta\sigma}\delta^{\tau\rho\lambda}_{\alpha\sigma\mu}
+{1\over2\kappa}\delta(N^\mu H_\mu{}^{\tau\rho})\right]dS_{\tau\rho}.
\end{equation}
(This result differs slightly from (\ref{varhamden}) because the
ADM
form of the Hamiltonian used here does not contain a term proportional
to $D\pi$.)
For example for the {\em Einstein} pseudotensor, use the Freud
superpotential (\ref{bh}) as
the Hamiltonian boundary term in (\ref{Ham}).  Then the boundary term
in the Hamiltonian variation has the integrand
$\delta(\pi^{\beta\sigma}N^\mu) \Gamma^\alpha{}_{\beta\lambda}
\delta^{\tau\rho\lambda}_{\alpha\sigma\mu}$,
which shows not only that  $\pi^{\beta\sigma}$
is to be held fixed on the boundary, but also that the appropriate
displacement vector field is $N^\mu=$ constant, and the reference
configuration here is Minkowski
space with a Cartesian reference frame.

A minor variation on the preceding analysis results from choosing a
superpotential with a contravariant index:
$H^{\mu\nu\lambda}\equiv H^{\mu[\nu\lambda]}$.
A further variation:
$
H^{\mu\nu\alpha}:=\partial_\beta H^{\mu\alpha\nu\beta}$,
 along with the symmetries
$H^{\mu\alpha\nu\beta}\equiv H^{\nu\beta\mu\alpha}\equiv
H^{[\mu\alpha][\nu\beta]}$ and $H^{\mu[\alpha\nu\beta]}\equiv0$,
leads to a {\em symmetric} pseudotensor---which then allows for a
simple definition of angular momentum, see \cite{MTW} \S20.2.  We can
cover these options simply by using the displacement vector
field to make modifications like
$N^\mu H_\mu{}^{\nu\lambda}\longrightarrow N_\mu H^\mu{}^{\nu\lambda}$.

In this way we see that each of the pseudotensors actually gives the
value of the quasilocal energy-momentum for an acceptable Hamiltonian.
In each case, via the Hamiltonian boundary variation symplectic
structure, this quasilocal energy-momentum is associated with some
definite physical boundary conditions \cite{CNC99}.  Note that this
same type of argument extends to superpotentials (i.e., Hamiltonian
boundary terms) that are more general than the classic
linear-in-displacement form associated with the traditional
pseudotensors.  In particular one can include first (and even higher)
derivatives of the displacement, as occurs in the Komar expression
(\ref{komar}).

In summary, similar to the Lagrangian analysis which we have
discussed, the boundary term in the Hamiltonian variation, ${\cal C}$,
generally does not vanish, so the Hamiltonian is not differentiable
for general field values.  A modification, achieved by adding a total
derivative term to the Hamiltonian, adjusts ${\cal B}$ without
changing the field equations and can compensate, making ${\cal C}$
vanish for suitable preselected boundary values.  The exact form of
such an adjustment still has infinite possibilities, this allows for
an infinite number of different gravitational energy definitions
\cite{Ber92b}.  However, each of them has its own unique expression
for the Hamiltonian boundary variation ${\cal C}$.  The symplectic
structure of this term reveals the implicit boundary conditions and
thereby gives a physical interpretation for each quasilocal
energy-momentum expression.  Thus, for each well-defined Hamiltonian
boundary expression, one can, via the Hamiltonian analysis, find its
associated symplectic structure which shows the built in control mode,
or equivalently, the implicit boundary conditions.

%%%%%%%%%%%%%%%%%%%%%%%%%%%%%%%%%%%%%%%%%%%%%%%%%%%%%%%%%%%%%%%%%%%%%%
%                                                                    %
%%%%%%%%%%%%%%%%%%%%%%%%%%%%%%%%%%%%%%%%%%%%%%%%%%%%%%%%%%%%%%%%%%%%%%
\section{Quasilocal Energy-Momentum}

Here we describe our Hamiltonian boundary term expressions for
quasilocal energy-momentum.  Our major tool is the symplectic analysis
of the Hamiltonian boundary variational principle.  We associate each
possible Hamiltonian boundary term expression with the boundary
conditions identified via the symplectic structure of the boundary
term in the variation of the Hamiltonian.  There are an infinite
number of possible Hamiltonian boundary terms, and correspondingly an
infinite number of possible boundary conditions.  We greatly reduce
this infinity by applying a {\it covariance} criteria.

In the previous section we saw that the Hilbert Hamiltonian
had problems asymptotically while the Einstein Hamiltonian gave good
asymptotic values but had a non-covariant response, being linear in the
connection.  These shortcomings necessitate, as we saw at the
Lagrangian level, the introduction of a reference geometry.
Hence for regulating the variational boundary term, a background
manifold with a suitable geometry,
$(\bar M, \bar g_{\mu\nu}, \bar \Gamma^\alpha{}_{\mu\nu})$, is
introduced as a reference
configuration.
The gravitational energy-momentum is understood to be measured with
respect to this selected background.
Any modification of the Lagrangian or Hamiltonian boundary term
changes the symplectic structure and the boundary
conditions.
Here, from the two examples we considered, we obtain
 modified versions
of ${\cal B}_H$ and ${\cal B}_E$ which have the same the control
modes,
respectively $\Gamma^\mu{}_{\alpha\beta}$ or $\pi^{\mu\nu}$ , but
their responses are given an improved ``covariant'' form.

For the metric density (in deference to the traditional choice of
variables we refer to it as the ``Dirichlet'') control mode, the
background is just what we need to make the responses become tensorial
objects without changing the control variables.  Its symplectic
structure in the variational boundary term is required to have the
form
\begin{equation}
{\cal C}_\pi^{\tau\rho}(N) =
      2 \Delta \Gamma^\mu{}_{\alpha\beta}       \,
      \delta \pi^{\beta\nu}\delta^{\alpha[\tau}_{\mu\nu} N^{\rho]},
\label{Cpi}
\end{equation}
where the $\Delta$ means the difference of variables between physical and
background configurations (i.e., $\Delta
\Gamma^\mu{}_{\alpha\beta}
:=   \Gamma^\mu{}_{\alpha\beta} - \bar \Gamma^\mu{}_{\alpha\beta}$
and $\Delta \pi^{\mu\nu}:= \pi^{\mu\nu}-\bar \pi^{\mu\nu}$).  Now the
response is a combination of $\Delta \Gamma$ which is a tensor.
Moreover the whole Hamiltonian boundary variation term is now the
projection along the displacement vector field of a four dimensionally
covariant object, a vector density which vanishes asymptotically
(spatially) with standard fall offs --- showing that the Hamiltonian
is differentiable on the space of all asymptotically flat fields.  In
order to obtain this desired ${\cal C}_\pi$, we must modify ${\cal
B}_E$.  The modified quasilocal energy-momentum boundary term, ${\cal
B}_\pi$, was found in \cite{Chen94,CNT95,CN99} to be
\begin{equation}
{\cal B}_\pi^{\mu\nu}(N) =
      N^\tau \pi^{\beta\lambda} \Delta \Gamma^\alpha{}_{\beta\gamma}
      \delta^{\mu\nu\gamma}_{\alpha\lambda\tau}
    + N^\tau \bar \Gamma^\alpha{}_{\beta\tau}
 \Delta \pi^{\beta\lambda}
   \delta^{\mu\nu}_{\alpha\lambda}. \label{Bpi1}
\end{equation}

Similarly,
for our connection (``Neumann'') control mode, the boundary term in the
Hamiltonian variation
can also be improved by incorporating reference quantities in the form
\begin{equation}
{\cal C}_\Gamma^{\tau\rho} =
    - 2 \Delta \pi^{\beta\nu}\, \delta \Gamma^\mu{}_{\alpha\beta}
 \delta^{\alpha[\tau}_{\mu\nu} N^{\rho]} .
\label{Cgamma}
\end{equation}
This symplectic expression is again the projection along the
displacement vector field of a four-covariant vector density which
automatically vanishes asymptotically (spatially, with standard fall
offs) indicating that the Hamiltonian is differentiable on the space
of all asymptotically flat fields.  This version follows
from the adjusted Hamiltonian boundary term \begin{equation}
{\cal B}_{\Gamma}^{\mu\nu} =
      N^\tau \bar \pi^{\beta\lambda} \Delta \Gamma^\alpha{}_{\beta\gamma}
      \delta^{\mu\nu\gamma}_{\alpha\lambda\tau}
    + N^\gamma \Gamma^\alpha{}_{\beta\gamma} \Delta \pi^{\beta\lambda}
      \delta^{\mu\nu}_{\alpha\lambda} \label{Bgamma1}.
\end{equation}
Note that the two modes are complimentary:  the Hamiltonian boundary
variation symplectic relation for one can be obtained from the other
just by interchanging the control-response roles.

From the variables at hand there are two other Hamiltonian boundary term
expressions which can be constructed:
\begin{eqnarray}
{\cal B}_0^{\mu\nu} &=&
      N^\tau \bar \pi^{\beta\lambda} \Delta \Gamma^\alpha{}_{\beta\gamma}
      \delta^{\mu\nu\gamma}_{\alpha\lambda\tau}
    + N^\gamma \bar\Gamma^\alpha{}_{\beta\gamma} \Delta \pi^{\beta\lambda}
      \delta^{\mu\nu}_{\alpha\lambda}, \\
{\cal B}_1^{\mu\nu} &=&
      N^\tau \pi^{\beta\lambda} \Delta \Gamma^\alpha{}_{\beta\gamma}
      \delta^{\mu\nu\gamma}_{\alpha\lambda\tau}
    + N^\gamma \Gamma^\alpha{}_{\beta\gamma} \Delta \pi^{\beta\lambda}
      \delta^{\mu\nu}_{\alpha\lambda}.
\end{eqnarray}
${\cal B}_0$ has the interesting property of being linear in
the dynamic variables $\pi$, $\Gamma$ while ${\cal B}_1$ is linear in
$\bar\pi$, $\bar\Gamma$.
The two associated Hamiltonian variation boundary terms (both of which
automatically vanish asymptotically with standard spatial fall offs) have a
remarkable $\Delta \leftrightarrow \delta$ symmetry:
 \begin{eqnarray}
{\cal C}_0^{\tau\rho} &=&
-\delta \Gamma^\alpha{}_{\beta\gamma}\Delta\pi^{\beta\sigma}N^\mu
\delta^{\tau\rho\gamma}_{\alpha\sigma\mu}-
\Delta\Gamma^\alpha{}_{\beta\gamma}N^\gamma\delta\pi^{\beta\sigma}
\delta^{\tau\rho}_{\alpha\sigma}, \\
{\cal C}_1^{\tau\rho} &=& \quad
\Delta \Gamma^\alpha{}_{\beta\gamma}\delta\pi^{\beta\sigma}N^\mu
\delta^{\tau\rho\gamma}_{\alpha\sigma\mu}+
\delta\Gamma^\alpha{}_{\beta\gamma}N^\gamma\Delta\pi^{\beta\sigma}
\delta^{\tau\rho}_{\alpha\sigma}.
\end{eqnarray}
However, neither has the $J^{[\tau}N^{\rho]}$ form of a projection
along $N^\mu$ of a 4-dimensionally covariant vector density, only
our two expressions (\ref{Bpi1},\ref{Bgamma1}) (or constant linear
combinations thereof) leading to (\ref{Cpi},\ref{Cgamma}) have this
desirable `covariant' property.

Returning to our two `covariant' expressions, there is a technical
hitch here that needs discussion.  Although the Hamiltonian variation
control-response symplectic structure has a nice covariant form, the
Hamiltonian boundary terms themselves (\ref{Bpi1},\ref{Bgamma1}) are
not fully covariant.  This is an inevitable consequence of our
particular style of first order `independent metric, frame and
connection' formulation, as we briefly explain here (the main
technical point is that we actually treat the connection as a one
form; for further remarks see appendix \ref{dynamic}).  The connection
is not a covariant object.  The Hamiltonian must generate the
evolution of the connection coefficients including the reference frame
gauge dependent part (which depends on the displacement vector field
differentially).  This latter task is the duty of the $N\Gamma D\pi$
term in the Hamiltonian.  The Hamiltonian boundary term then includes
an associated piece with the form $N \Gamma\Delta\pi$.  The
contribution of this piece to the value of the energy-momentum is a
mixture of a covariant physical contribution along with an
energy-momentum associated with the particular reference frame.
Fortunately these contributions can easily be separated by using the
identity
\begin{equation}
N^\mu \Gamma^\alpha{}_{\beta\mu}\equiv D_\beta N^\alpha-\partial_\beta
N^\alpha,
\end{equation}
to replace the $N\Gamma$ terms.  The $\partial N$ terms produce a
noncovariant (reference frame dependent) unphysical contribution
(which can usually be made to vanish in a specially selected frame)
and should be dropped (for the purposes of calculating physical
energy-momentum but not for calculating the evolution equations).
This leads to the final {\em fully covariant} form of our Hamiltonian
boundary quasilocal energy-momentum expressions:
\begin{eqnarray}
{\cal B}_\pi^{\mu\nu}(N) &=&
      N^\tau \pi^{\beta\lambda} \Delta \Gamma^\alpha{}_{\beta\gamma}
      \delta^{\mu\nu\gamma}_{\alpha\lambda\tau}
    + \bar D_\beta N^\alpha \Delta \pi^{\beta\lambda}
      \delta^{\mu\nu}_{\alpha\lambda}, \label{Bpi} \\
{\cal B}_{\Gamma}^{\mu\nu}(N) &=&
      N^\tau \bar \pi^{\beta\lambda} \Delta \Gamma^\alpha{}_{\beta\gamma}
      \delta^{\mu\nu\gamma}_{\alpha\lambda\tau}
    + D_\beta N^\alpha \Delta \pi^{\beta\lambda}
      \delta^{\mu\nu}_{\alpha\lambda}. \label{Bgamma}
\end{eqnarray}
We wish to emphasize that
an alternate, {\em fully covariant, direct} derivation of these
expressions
can be obtained from a different representation as indicated in
Appendix \ref{dynamic}.

After a bi-metric manipulation (see Appendix \ref{BIM}), the above
expressions can be rewritten in the following compact and remarkably
similar forms:
 \begin{eqnarray}
{\cal B}_\pi^{\mu\nu} &=& 2 \Delta(\pi^{\lambda[\nu} D_\lambda N^{\mu]})
    + N^{\nu} k^{\mu}(\pi)-N^{\mu} k^{\nu}(\pi), \label{Bk} \\
{\cal B}_\Gamma^{\mu\nu} &=&
      2\Delta(\pi^{\lambda[\nu} D_\lambda N^{\mu]})
    + N^{\nu} k^{\mu}(\bar\pi)-N^{\mu} k^{\nu}(\bar\pi),
     \label{Bl}
\end{eqnarray}
where
\begin{equation}
k^\mu(\pi) := \pi^{\mu\nu} \Delta\Gamma^\lambda{}_{\nu\lambda}
    - \pi^{\alpha\beta} \Delta\Gamma^\mu{}_{\alpha\beta}, \label{kk}
\end{equation}
and $
k^\mu(\bar\pi)$ has the same form with $\bar\pi$ replacing $\pi$.

At first we were surprised to learn that our Dirichlet expression
(\ref{Bpi}) is {\it exactly} identical with an expression obtained by
Katz {\sl et al.} \cite{LBKB95,KBLB97,KL97} which was derived in a
completely different way, namely by applying the Noether conservation
theorem to the Lagrangian density (compare with
(\ref{IE})):
 \begin{eqnarray}
{\cal L}_\pi &=&
      \pi^{\mu\nu} R_{\mu\nu} + \partial_\mu k^\mu(\pi)
    - \bar \pi^{\mu\nu} \bar R_{\mu\nu} \nonumber \\
   &=& - \pi^{\mu\nu}
      (\Delta\Gamma^\lambda{}_{\rho\lambda} \Delta\Gamma^\rho{}_{\mu\nu}
    - \Delta\Gamma^\lambda{}_{\mu\rho} \Delta\Gamma^\rho{}_{\nu\lambda})
    +  \Delta \pi^{\mu\nu} \bar R_{\mu\nu},
\end{eqnarray}
which includes background terms in addition to terms quadratic in the
first derivatives of $g_{\mu\nu}$.  In retrospect we realize that our
having found an identical energy-momentum expression is not so
surprising after all.  The Hamiltonian approach, as we discussed, is
closely connected with the Noether approach.  Moreover our covariance
requirement leaves little room in the Hamiltonian boundary term for
anything else except expressions that can be inherited from a suitable
four dimensionally covariant Lagrangian.

Comparing the remarkable similarity in the form of our alternate
Neumann expression (\ref{Bgamma}) with that of (\ref{Bpi}),
invites us
to consider also obtaining it from a Lagrangian density.  The desired
result is obtained simply by interchanging the roles of $g$ and $\bar
g$ (consequently $\Delta\to-\Delta$) followed by an overall sign
change.  Thus, we found that (\ref{Bl}) can be derived by the same
Noether argument used by Katz and coworkers from the following
Lagrangian, which is quadratic in the first derivatives of $\bar
g_{\mu\nu}$:
 \begin{eqnarray}
{\cal L}_{\Gamma} &=&
      \pi^{\mu\nu} R_{\mu\nu} + \partial_\mu k^\mu(\bar\pi)
    - \bar \pi^{\mu\nu} \bar R_{\mu\nu} \nonumber \\
   &=& - \bar \pi^{\mu\nu}
      (\Delta\Gamma^\lambda{}_{\rho\lambda} \Delta\Gamma^\rho{}_{\mu\nu}
    - \Delta\Gamma^\lambda{}_{\mu\rho} \Delta\Gamma^\rho{}_{\nu\lambda})
    +  \Delta \pi^{\mu\nu} R_{\mu\nu}.
\end{eqnarray}

Via the Hamiltonian boundary term symplectic structure we identified
the two quasilocal energy expressions (\ref{Bpi},\ref{Bgamma}) as
corresponding to Dirichlet or Neumann type boundary conditions
respectively.  Comparing their respective Lagrangians we see the
relation between these two expressions from a new point of view.
There is an amazing symmetry relating the two expressions:  one passes
into the other simply by interchanging the role of the dynamic
physical and background variables.  Hence we can regard our
``Neumann'' expression (\ref{Bgamma}) as giving the energy-momentum of
the ``reference'' geometry measured with respect to the ``dynamic''
geometry using ``Dirichlet'' boundary conditions.  And, likewise, our
``Dirichlet'' expression gives the ``Neumann'' energy-momentum for the
``reference'' geometry compared to the ``dynamic'' geometry.  This
symmetry, however, has an intriguing asymmetry:  one might have
expected the energy-momentum of the dynamic space referenced to the
background to have the same magnitude when the roles are reversed {\em
without} any reversal in the type of boundary condition (on the other
hand, one could argue that asymmetries are common when the reference
point is changed, e.g., going from 4 to 5 and back to 4 can be
described as a 25\% increase followed by a 20\% decrease).  We suspect
that there is some, as yet unidentified, underlying principle which
could have been used to anticipate this curious symmetry and
asymmetry.

Our two boundary expressions are not the only ones for gravitational
energy-momentum.  They are simply the only ones which satisfy our
covariant Hamiltonian symplectic structure criterion.  Covariance is a
very important property; we believe that a covariant theory should
have covariant quasilocal energy-momentum.  Nevertheless it is well to
keep in mind that some other property may be regarded as even more
desirable (also, perhaps our particular implementation of the
covariance requirement could be generalized).  Then one could exploit
the freedom in selecting the Hamiltonian boundary term to achieve a
different goal.  For example Kijowski and Jezierski \cite{JK90,Kij97}
have used the constraints and the actual boundary conditions required
by the field equations to identify and control certain variables
representing the true physical degrees of freedom.  This necessitates
decomposing the dynamic fields into various space, time and boundary
components.  Consequently their expression for quasilocal
energy-momentum is not covariant (although it would be interesting to
try to recast it into that form).  More recently, following along the
lines of Rosenfeld and Belinfante, Petrov and Katz have used what
amounts to the same Hamiltonian boundary term freedom that we have
exploited to achieve
a ``symmetric'' energy-momentum expression \cite{PK99}.  To any such
alternate expressions one can apply, just as we did for the
pseudotensors, our Hamiltonian boundary variation symplectic analysis
to reveal the implicit spatial boundary conditions.

\section{Conclusions}

In summary,
variational principles can yield not only field equations but also
boundary conditions; the latter can be modified by adding a total
derivative, which is equivalent to a boundary term.  The Hamiltonian
boundary term for dynamic spacetime governs the value of the
Hamiltonian.  For each finite region its value yields a quasilocal
energy-momentum.  The boundary term in the variation of the
Hamiltonian has a symplectic structure, which is uniquely determined
by the choice of quasilocal expression.  Requiring it to vanish
associates to each different quasilocal expression distinct boundary
conditions.  This approach provides a physical interpretation for many
of the well-known gravitational energy-momentum expressions including
all of the pseudotensors, associating each with unique boundary
conditions.  Among the infinite possibilities, we found only two
Hamiltonian-boundary-term quasilocal expressions which correspond to
{\em covariant} boundary conditions; they are respectively of the
Dirichlet or Neumann type.  Our Dirichlet expression coincides with
the expression recently obtained by Katz and coworkers using Noether
arguments and a fixed background.  A modification of their argument
yields our Neumann expression.

Some key points we have noted in our analysis are:

\begin{itemize}

\item The boundary-variation-symplectic-structure principle connects
the choice of boundary term with boundary conditions.
The Lagrangian boundary term can be adjusted to affect a
canonical transformation.  It governs the boundary conditions
on the 3-dimensional boundary of a spacetime region, including the
initial time spacelike hypersurface.

\item The Hamiltonian boundary term governs the boundary conditions on
the 2-dimensional boundary of the spatial region at each instant of
time.  The value of the Hamiltonian for dynamic geometry theories
including general relativity is determined entirely by the Hamiltonian
boundary term.  It gives the quasilocal energy-momentum.  Our freedom
to adjust the Hamiltonian boundary term is justified by the
conservation law.  The Hamiltonian boundary term depends on the
displacement vector, hence it has (in principle) more freedom than is
available at the Lagrangian level.  However the value of the
Hamiltonian, the energy-momentum, also has physical `correspondence
limit' constraints which have no analog for the Lagrangian.  The
boundary term freedom we exploit here is essentially the same freedom
used in constructing `new improved symmetric energy-momentum tensors'.

\item The Einstein and M{\o}ller (Komar) pseudotensors arise quite
naturally, but the latter has several more shortcomings.
All of the pseudotensor superpotentials are possible Hamiltonian
boundary terms.  Consequently all pseudotensors have quasilocal
energy-momentum which is identical to the value of the Hamiltonian for
an acceptable choice of boundary term,  which, in turn, corresponds to some
definite boundary conditions.

\item Our `covariance' criterion removes most of the freedom (leaving
only two choices).  In
hindsight we see that it essentially restricts us to expressions which
could be obtained (without any adjustments by hand) by projecting
a judicious choice of Lagrangian boundary term.
 Our Dirichlet mode unexpectedly coincides with that of Katz et
al.  In retrospect that is no surprise.
 Our Neumann mode can be interpreted as the Katz et al.
energy-momentum
of the reference geometry referred to the dynamic geometry.

\end{itemize}

 We note some features of our quasilocal
energy-momentum expressions:

\begin{itemize}

\item We found only two ``covariant'' Hamiltonian boundary
expressions.  They each give rise to a boundary term in the variation
of the Hamiltonian which has the form of a projection of a covariant
vector density along the displacement vector field.  The form of this
variation boundary term shows that the respective Hamiltonians evolve
field values with Dirichlet or Neumann type boundary conditions.  With
standard fall offs, the two Hamiltonians have well defined variational
derivatives on the space of asymptotically flat fields at spatial
infinity.

\item Our expressions depend on a reference configuration, which is
required only on the boundary.  The reference configuration
determines the zero point for all of the quasilocal quantities.
The obvious choice is Minkowski space; alternatives
which may be  more appropriate for certain applications include
(anti-)de Sitter space, a Friedmann-Robertson-Walker cosmology, and
Schwartzschild geometry.  Some options for attaching
an appropriate reference configuration to a dynamic boundary were
discussed in \cite{CN99}.

\item Our expressions also depend on a displacement vector field which
selects the associated component of the quasilocal energy-momentum.
How to choose the exact form of this vector field was discussed in
\cite{CN99}; the recommended choice is a Killing vector of the
reference geometry.  In addition to energy-momentum (obtained from a
spacetime translation), for a suitable choice of rotational
displacement, the expressions also give angular momentum.

\item Our expressions reduce to expressions proposed by others in the
appropriate limits, in particular to the well known quasilocal
expressions of Brown \& York \cite{BY93} and asymptotically to that of
Beig \& {\sc \'o} Murchadha \cite{BO87}.  Asymptotically they are
equivalent to an expression which gives the expected values at spatial
infinity (for asymptotically flat {\em and} anti-de Sitter solutions)
\cite{HN93}.  Moreover, asymptotically, at future null infinity, our
Dirichlet expression yields the expected Bondi values \cite{HN96}.
Quasilocally, we have evaluated them for spherically symmetric
spacetimes \cite{CNT95,CN99}.

\item Katz and coworkers have applied their expression (equivalent to
our Dirichlet expression) at future null infinity \cite{KL97} to
cosmology \cite{KBLB97} and Mach's principle \cite{LBKB95}.  We have
applied our formulation to black hole thermodynamics \cite{CNT95,CN99}
to obtain the first law and an expression for the entropy.

\end{itemize}

More generally our work reveals some of the merits of the symplectic
Hamiltonian boundary variational principle.  In particular it allows
us to supplement the usual (correspondence limit to weak field and
asymptotic forms) constraints on quasilocal energy-momentum
expressions with a principle which connects each quasilocal expression
with a distinct boundary condition.  Coupled with the covariance
criteria the form of the quasilocal energy-momentum expression is then
strongly restricted.

\section*{Acknowledgments}
We are very grateful to R.S. Tung, A.N. Petrov and J. Katz for useful
discussions and correspondence.
The work was supported by the National Science Council of the Republic
of China under grants No. NSC89-2112-M-008-020 and
NSC89-2112-M-008-016.

%%%%%%%%%%%%%%%%%%%%%%%%%%%%%%%%%%%%%%%%%%%%%%%%%%%%%%%%%%%%%%%%%%%%%%
% Appendix                                                           %
%%%%%%%%%%%%%%%%%%%%%%%%%%%%%%%%%%%%%%%%%%%%%%%%%%%%%%%%%%%%%%%%%%%%%%
\begin{appendix}
\section{Dynamical Details} \label{dynamic}
Our Hamiltonian formalism is adapted to evolving the components of
objects including the connection coefficients, hence it includes some
(unphysical) dynamic reference frame gauge generation features.  There
are alternate representations (e.g., \cite{Kij97}) in which these
terms do not show up.

Our general approach is to work with a dynamical Lagrangian and
Hamiltonian formulation which gives independent equations for evolving
the frame, metric and connection, and which handles a wide range of
theories, including geometric gravity theories and gauge theories in a
uniform way \cite{CN99}.  Note that in such a formalism the
Hamiltonian
must include the ability to generate a general time dependent (purely
gauge) evolution of the frame and the associated induced effects on
the components of geometric objects.  In general we found that there
are certain technical advantages in using a differential form
representation.  However in the present work we wanted to make our
results for the specific case of Einstein's GR more accessible to
others, so we transcribed it into the ordinary ``holonomic frame''
representation.  To achieve this some choices must be made:  in
particular, how to deal with the metric, frame and connection
variables and how to impose the vanishing torsion and metric
compatible constraints.  We want to keep our first order form, so we
certainly need the connection and metric to be independent at least to
some extent.  We elected to impose vanishing torsion `a priori' and
thus to use a symmetric connection and a variational principle which
would give the metric compatible condition as a (vacuum) field
equation.
Because we are using a holonomic frame the evolution of the frame is
rather trivial, so we dropped it and its conjugate momentum from our
list of dynamic variables.  Nevertheless we did not want to depart far
from the form of our earlier more general work.  Thus the expressions
given here are actually obtained by specializing our earlier work.
In particular our time derivative is specified by projecting the Lie
derivative $\Lie_N:=di_N+i_Nd$ of {\em components} of the connection
one-form and its conjugate momentum 2-form:
  \begin{eqnarray}
\dot\Gamma{}^\alpha{}_{\beta\gamma} dx^\gamma
&:=&\Lie_N(\Gamma{}^\alpha{}_{\beta\lambda}dx^\lambda) =
\big(N^\mu\partial_\mu\Gamma^\alpha{}_{\beta\gamma}+
\Gamma^\alpha{}_{\beta\mu}\partial_\gamma N^\mu\big)dx^\gamma, \\
\dot\pi^{\beta\sigma}\epsilon_{\alpha\sigma}
&:=&\Lie_N(\pi^{\beta\sigma}\epsilon_{\alpha\sigma})
=\big(N^\lambda\partial_\lambda\pi^{\beta\sigma} \delta^\rho_\alpha
+{1\over2}\pi^{\beta\nu}\partial_\mu N^\lambda
\delta^{\mu\rho\sigma}_{\alpha\nu\lambda}\big)
\epsilon_{\rho\sigma},
 \end{eqnarray}
 where
 $\epsilon_{\mu\nu}:=
(1/2)\epsilon_{\mu\nu\alpha\beta}dx^\alpha\wedge dx^\beta$.
  These
``definitions'' differ from the usual holonomic expression of the
components of the Lie derivative for the contravariant metric density:
\begin{equation}
 \Lie_N\pi^{\mu\nu}:=\partial_\lambda( N^\lambda
\pi^{\mu\nu}) -\pi^{\alpha\nu}\partial_\alpha N^\mu
-\pi^{\mu\alpha}\partial_\alpha N^\nu
\label{LieNpi}
 \end{equation}
 and the connection coefficients:
 \begin{equation} \Lie_N
\Gamma^\alpha{}_{\beta\gamma}:=-R^\alpha{}_{\beta\gamma\mu}N^\mu
+D_\gamma D_\beta N^\alpha.
\label{LieNGamma}
 \end{equation}
 The difference, however, shows up
only in terms proportional to the derivative of N.  (All such terms
vanish if the coordinates are adapted so that $\Lie_N$ reduces to
$\partial_t$).  A feature of our approach is a frame-gauge
generating term in the Hamiltonian density and an associated term in
the Hamiltonian boundary quasilocal expression.  We like this set up
for it is just the way things come out for the vector potential (i.e.
connection one-form) in gauge theories like electromagnetism and
Yang-Mills.

It is certainly possible to use the usual Lie derivative in a
Hamiltonian formulation (see, e.g., \cite{Chr,KT79,Kij97}).  We choose
to avoid it also because it includes an inconvenient for us
second derivative of $N$ (which necessitates adjustments in our
argument regarding the form of and the vanishing of the Hamiltonian
density).  A price we pay is then an awkward term like
$N^\tau\bar\Gamma^\alpha{}_{\beta\tau}
\Delta\pi^{\beta\lambda}\delta^{\mu\nu}_{\alpha\lambda}$ in each of
our quasilocal energy-momentum Hamiltonian boundary expressions.  We
have argued that such terms are necessary to give us the Hamiltonian
evolution and boundary variation symplectic structure in our
representation.  Unfortunately the quasilocal energy-momentum, defined
as the value of our Hamiltonian, consequently includes both a physical
and an unphysical, reference frame gauge dependent, contribution.  To
separate these effects we rearrange the symmetric connection identity
$
(\Lie_N e_\beta)^\alpha\equiv [N,e_\beta]^\alpha \equiv
\big(\nabla_N e_\beta-\nabla_\beta N\big)^\alpha $
to give
\begin{equation}
N^\mu \Gamma^\alpha{}_{\beta\mu} \equiv D_\beta N^\alpha+(\Lie_N
e_\beta)^\alpha,
 \end{equation}
which can be used to replace the $N\Gamma$ factors.  The $\Lie_N
e_\beta$ term is a non-covariant, dynamic reference frame piece.  Its
contribution to the energy-momentum can be thought of as an energy
associated with the observer.  In fact, for any given displacement
vector field $N$, we can choose the reference frame $e_\beta$ so that
it vanishes.

Having introduced this identity, an alternative approach is available.
We could treat this term in the same way as its analogue is dealt with
in other representations, in particular Kijowski's \cite{Kij97}.  Note
that, since it includes a time derivative, it really has no place in a
Hamiltonian.  Rather it should be treated as term belonging to the
$p_k\dot q^k$ part of the action, a term that shows up in a
2-dimensional integral over the boundary of the spacelike hypersurface
rather than in the 3-dimensional hypersurface integral.  An easy way
to establish the association between this part of our representation
and Kijowski's is to consider the frame to be orthonormal.  Then its
time evolution is just an instantaneous Lorentz boost (in the
spacetime
2-plane orthogonal to the spatial boundary) by a hyperbolic angle
$\dot \alpha \delta t$.  The associated `conjugate momentum' is the
area of the 2-surface.  Hayward \cite{Hay93} gives another route to
time derivative terms on the spatial 2-boundary.  He uses the fact
that the total boundary term in the Einstein action (\ref{LE}) can be
expressed as the extrinsic curvature of the boundary.  The standard
definition of the extrinsic curvature involves the normal to the
boundary surface.  But converting the total derivative form to a
surface integral is then a delicate task, as the normal is
discontinuous on the corners of the usual 3-boundary, which consists
of an initial and final constant time spacelike hypersurface connected
by a topologically $S^2\times [t_i,t_f]$ type 3-manifold.  This leads
to contributions in the action given by the difference between an
integral over the final and initial 2-boundary.  Contributions which
can, in turn, be written as the integral over time of a total time
derivative of a 2-boundary term.

Actually it is not difficult to obtain a fully covariant Hamiltonian
density with our fully covariant quasilocal boundary terms.
Beginning from the Hilbert Lagrangian (\ref{LH}), the Hamiltonian can
be derived by using the usual Lie derivative of a
connection (\ref{LieNGamma})
 \begin{eqnarray}
{\cal H}^\mu(N) &:=& \Lie_N \Gamma^\alpha{}_{\beta\nu}
\pi^{\beta\sigma}
     \delta^{\mu\nu}_{\alpha\sigma} - N^\mu {\cal L}_H \\
   &\equiv& -\frac12 N^\nu R^\alpha{}_{\beta\gamma\lambda}
\pi^{\mu\beta}       \delta^{\mu\lambda\gamma}_{\alpha\nu\sigma}
    - D_\beta N^\alpha \delta^{\mu\nu}_{\alpha\sigma}
     D_\nu \pi^{\beta\sigma}
    + \partial_\nu \left( D_\beta N^\alpha \pi^{\beta\sigma}
      \delta^{\mu\nu}_{\alpha\sigma} \right).
\end{eqnarray}
The boundary term here is just the Komar superpotential (with the
normalization that gives half of the desired energy-momentum).  The
boundary term in the variation of this Hamiltonian will not
automatically vanish asymptotically; hence this Hamiltonian requires
the explicit functional form of the connection to be fixed on the
boundary even asymptotically.  Consequently this Hamiltonian should be
adjusted.  Replacing the boundary term by one of the improved boundary
terms (\ref{Bpi}) or (\ref{Bgamma}) gives a fully 4-covariant
Hamiltonian for general relativity.  Explicitly calculating the
resultant boundary term in the variation of the Hamiltonian then leads
to the desirable asymptotically well behaved covariant symplectic
structures (\ref{Cpi}), (\ref{Cgamma}).  For constant components
$N^\mu$ these fully covariant Hamiltonian density plus boundary term
expressions reduce to
(\ref{hamden},\ref{BH},\ref{Bpi1},\ref{Bgamma1}).

 \section{Geometry of Bi-metric Spacetime} \label{BIM}
A background is needed to determine well-defined
conserved quantities
in
GR. For the special choice of mapping and coordinates such that a point
$P$ of the physical configuration is mapped into a point $\bar P$ of the
background and both are given the same coordinates $x^\mu$, the whole system
can be looked at as a spacetime $M$ possessing two metrics $g_{\mu\nu}$
and $\bar g_{\mu\nu}$.
  Geometric quantities with respect to each
metric can then be reformulated in terms of the difference between them.
In particular each metric determines its own
associated Levi-Civita
connection and Riemannian geometry.

The simplest case for the connection, from which all others can be derived is
\begin{equation}
\left( D_\mu - \bar D_\mu \right) N^\alpha
    = \Delta \Gamma^\alpha{}_{\mu\nu} N^\nu,
\end{equation}
where the variables and operators are denoted with or without a bar
consistently with the notation for the metric, and the symbol $\Delta$
means the difference of operands between two metrics such as
$\Delta \Gamma = \Gamma - \bar\Gamma$.
This identity shows that $\Delta \Gamma$, being the difference between two
connections is a covariant tensorial object.

The Ricci tensor $R_{\mu\nu}$ ($\bar R_{\mu\nu}$) with respect to
$\Gamma^\alpha_{\beta\mu}$ ($\bar \Gamma^\alpha_{\beta\mu}$) can be
rewritten, respectively as
\begin{eqnarray}
 R_{\mu\nu} &=&
 \bar D_\lambda \Delta\Gamma^\lambda{}_{\mu\nu}
 - \bar D_\mu \Delta\Gamma^\lambda{}_{\nu\lambda}
 + \Delta\Gamma^\rho{}_{\mu\nu} \Delta\Gamma^\lambda{}_{\rho\lambda}
 - \Delta\Gamma^\rho{}_{\mu\lambda} \Delta\Gamma^\lambda{}_{\nu\rho} +
\bar R_{\mu\nu}, \\
 \bar R_{\mu\nu} &=&
 - D_\lambda
\Delta\Gamma^\lambda{}_{\mu\nu} + D_\mu \Delta\Gamma^\lambda{}_{\nu\lambda}
 + \Delta\Gamma^\rho{}_{\mu\nu} \Delta\Gamma^\lambda{}_{\rho\lambda} -
\Delta\Gamma^\rho{}_{\mu\lambda} \Delta\Gamma^\lambda{}_{\nu\rho} + R_{\mu\nu}.
\end{eqnarray}

Two other useful identities concern the total derivative terms, which
are added to the Hilbert Lagrangian density in order to make the
Lagrangian density quadratic in the first order derivatives of the metric:
\begin{eqnarray}
\partial_\mu k^\mu(\pi) &=& - \pi^{\mu\nu} \Bigl\{
      ( \bar D_\lambda \Delta\Gamma^\lambda{}_{\mu\nu}
    - \bar D_\mu \Delta\Gamma^\lambda{}_{\nu\lambda} )
    + 2( \Delta\Gamma^\rho{}_{\mu\nu} \Delta\Gamma^\lambda{}_{\rho\lambda}
    - \Delta\Gamma^\rho{}_{\mu\lambda} \Delta\Gamma^\lambda{}_{\nu\rho} )
      \Bigr\}, \\
\partial_\mu k^\mu(\bar\pi) &=& - \bar \pi^{\mu\nu}
      ( \bar D_\lambda \Delta\Gamma^\lambda{}_{\mu\nu}
    - \bar D_\mu \Delta\Gamma^\lambda{}_{\nu\lambda} ),
\end{eqnarray}
where the $k^\mu(\pi)$ and $k^\mu(\bar\pi)$ were defined in connection
with (\ref{kk}).
 \end{appendix}

%%%%%%%%%%%%%%%%%%%%%%%%%%%%%%%%%%%%%%%%%%%%%%%%%%%%%%%%%%%%%%%%%%%%%%
% References                                                         %
%%%%%%%%%%%%%%%%%%%%%%%%%%%%%%%%%%%%%%%%%%%%%%%%%%%%%%%%%%%%%%%%%%%%%%

%\end{multicols}  %%%%
\end{document}